\documentclass[runningheads]{llncs}
\usepackage[T1]{fontenc}
\usepackage{graphicx}
\usepackage{amsmath}
\usepackage{multirow}
\usepackage{booktabs}
\usepackage{float}
\usepackage{placeins}
\usepackage{cite}
\usepackage{hyperref}
\usepackage{academicons}
\usepackage{enumitem}
\usepackage{sectsty}
\usepackage{cleveref}
\usepackage{textcomp}
\usepackage{marvosym}
\usepackage{scalerel}
\usepackage{listings}
\usepackage{pifont}
\usepackage{makecell}
\usepackage{array}
\usepackage{amsmath}
\usepackage{graphicx}
\usepackage{booktabs}
\usepackage{resizegather}
\usepackage{caption}
\usepackage{siunitx}
\usepackage{xcolor}
\usepackage{amsfonts}

\hypersetup{
  colorlinks=true,
  linkcolor=blue,
  citecolor=blue,
  urlcolor=blue
}

\usepackage{url}

\begin{document}
\setlength{\abovedisplayskip}{5pt}
\setlength{\belowdisplayskip}{5pt}

\title{\textbf{SLIE: A Secure and Lightweight Cryptosystem for Data Sharing in IoT Healthcare Services\thanks{\textit{ Paper has been accepted for publication in the Proceedings of the
23th International Conference on Service-Oriented Computing
(ICSOC 2025, \url{https://icsoc2025.hit.edu.cn/})}}}}

\titlerunning{SLIE: Secure IoT Healthcare Services Communication}

% \textsuperscript{(\Letter)}
\author{
Ha Xuan Son\inst{1}\href{https://orcid.org/0000-0002-5336-0566}{\includegraphics[scale=0.004]{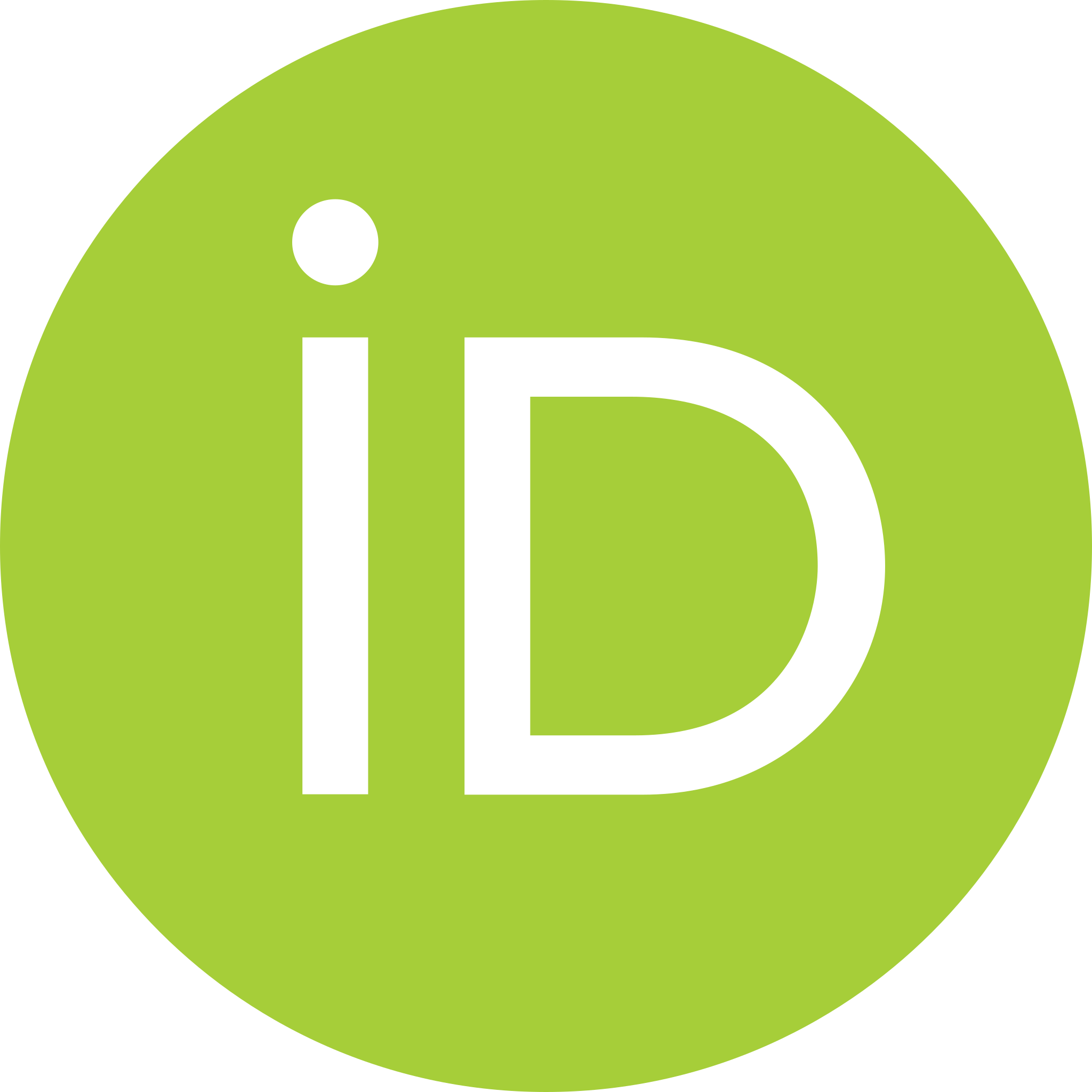}}
Nguyen Quoc Anh\inst{2}\href{https://orcid.org/0009-0008-0702-743X}{\includegraphics[scale=0.004]{sugg.png}} \and
Phat T. Tran-Truong \inst{3,4}\thanks{Corresponding author.}\href{https://orcid.org/0000-0003-3199-6333}{\includegraphics[scale=0.004]{sugg.png}} \and
Le Thanh Tuan\inst{5} \and
Pham Thanh Nghiem\inst{6}
}

\authorrunning{H. X. Son et al.}

\institute{
RMIT University, Ho Chi Minh City, Vietnam \\
\email{ha.son@rmit.edu.vn} \and
Hitachi Digital Services, Ho Chi Minh City, Vietnam \\
\email{anh.quocnguyen@hitachids.com} \and
Faculty of Computer Science and Engineering, Ho Chi Minh City University of Technology (HCMUT), Ho Chi Minh City, Vietnam \and Vietnam National University Ho Chi Minh City (VNU-HCM), Ho Chi Minh City, Vietnam \\
\textsuperscript{$\star$}\email{phatttt@hcmut.edu.vn}  \and NLS Tech Vietnam, Ho Chi Minh City, Vietnam \\
\email{tuan@nlstech.net} \and
Polarista A.I, Ho Chi Minh City, Vietnam \\
\email{nghiem.pham@polarista.vn}
}

\maketitle
\begin{abstract}
The Internet of Medical Things (IoMT) has revolutionized healthcare by transforming medical operations into standardized, interoperable services. However, this service-oriented model introduces significant security vulnerabilities in device management and communication, which are especially critical given the sensitivity of medical data. To address these risks, this paper proposes SLIE (Secure and Lightweight Identity Encryption), a novel cryptosystem based on Wildcard Key Derivation Identity-Based Encryption (WKD-IBE). SLIE ensures scalable trust and secure omnidirectional communication through end-to-end encryption, hierarchical access control, and a lightweight key management system designed for resource-constrained devices. It incorporates constant-time operations, memory obfuscation, and expiry-based key revocation to counter side-channel, man-in-the-middle, and unauthorized access attacks, thereby ensuring compliance with standards like HIPAA and GDPR. Evaluations show that SLIE significantly outperforms RSA, with encryption and decryption times of 0.936ms and 0.217ms for 1KB of data, an 84.54\% improvement in encryption speed, a 99.70\% improvement in decryption speed, and an energy efficiency of \SI{0.014}{\joule}/KB.

% Security tests show 0\% MITM, $<5\%$ side-channel, and $<5\%$ power analysis success. Evaluations demonstrate 0.936ms encryption and 0.217ms decryption for 1KB data, with \SI{0.014}{\joule} per \SI{1024}{\byte} energy efficiency, surpassing RSA by 84.54\% in encryption and 99.70\% in decryption time, offering key-creation mechanism over ChaCha20. NIST-aligned hybrid encryption ensures post-quantum readiness, posing SLIE as a versatile cryptosystem.

\keywords{WKD-IBE Encryption \and IoT Services \and Healthcare Services \and IoT Security \and Data Sharing}
\end{abstract}

\section{Introduction}
\label{intro}
The growing diversity of IoMT devices has enabled novel telehealth and telemedicine business models through the servitization of medical operations within service-oriented architectures. These architectures facilitate remote healthcare delivery by dynamically composing services to maximize value for stakeholders, including providers and patients \cite{bouguettaya2021internet}. By transforming IoT devices into IoT service providers, the IoMT ecosystem enables crowdsourced functionality, where devices not only collect data but also share resources and coordinate actuation across distinguished networks.

Despite these advancements, healthcare's unique security challenges present critical adoption barriers. Medical datasets contain 50\% more sensitive information than other sectors, making them high-priority targets for cyberattacks. This is reflected in a 45\% increase in ransomware incidents and an average breach cost of \$10.93 million \cite{in8, in10}. While centralized systems like Electronic Health Records (EHRs) and cloud platforms enhance accessibility, their API-dependent architectures introduce vulnerabilities to unauthorized access \cite{in11}. Existing cryptographic solutions, though promising, often fail to address the fundamental heterogeneity of IoMT ecosystems, which encompass devices ranging from wearables to implantables with disparate energy, storage, and computational capabilities \cite{bouguettaya2017service}. This diversity necessitates adaptive service models that holistically address both functional (e.g., data acquisition) and non-functional (e.g., real-time performance, security) requirements.

% Asymmetric alternatives improve key management but incur computational overhead. Hybrid approaches often prove impractical for real-time IoMT scenarios \cite{a9}. 
Current security solutions remain inadequate as they either: (1) address communication security in isolation, or (2) fail to accommodate the heterogeneous nature of IoMT devices. While symmetric encryption (e.g., AES) offers efficiency, it suffers from key distribution vulnerabilities \cite{in15}. Emerging solutions based on Attribute-Based Encryption (ABE), Hardware Security Modules, Trusted Execution Environments, and Blockchain provide promising directions for decentralized key management, but require careful adaptation to IoMT's unique constraints of heterogeneous devices and strict regulatory requirements \cite{in21}. To overcome these limitations, we propose SLIE (Secure and Lightweight Identity Encryption) - a novel cryptosystem optimized for omnidirectional IoMT service communication. Our proposed cryptosystem makes three fundamental advances in IoMT security:
\begin{itemize}
    \item \textbf{Hierarchical Key Management:} A novel WKD-IBE framework that enables efficient, fine-grained access control across distributed IoMT devices while maintaining minimal computational overhead.
    \item \textbf{Resource-Aware Access Control:} Non-commutative authorization mechanisms specifically optimized for constrained devices, ensuring secure communication without compromising performance.
    % \item \textbf{Side-Channel Resilience:} Dynamic protection against physical attacks through three synergistic techniques: (1) Constant-time cryptographic operations; (2) Memory access obfuscation; (3) Randomized execution timing
    \item \textbf{Automated Compliance:} An expiry-based key revocation that enforces HIPAA/GDPR requirements while minimizing management overhead.
\end{itemize}

\begin{figure}[!htb]
    \centering
    \includegraphics[width=0.7\textwidth]{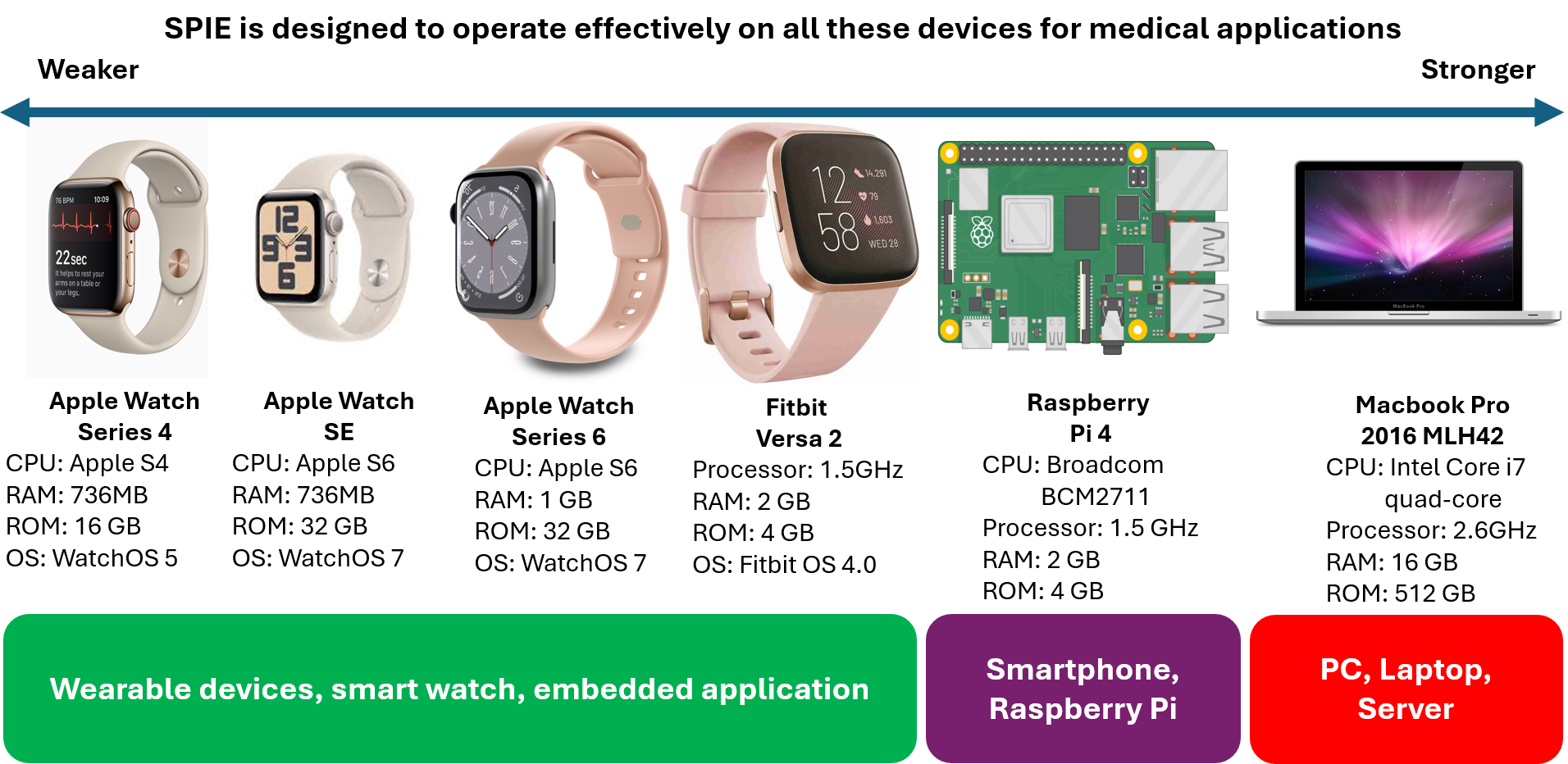}
    \caption{SLIE compatibility spectrum across healthcare IoT devices}
    \label{fig:evaluation_scenario}
\end{figure}

Our comprehensive experiments demonstrate SLIE's robust performance across diverse IoMT device capabilities (see \autoref{fig:evaluation_scenario}), delivering three key security advantages: (1) end-to-end encryption with proven resistance against MITM attacks through DBDH-secure cryptographic primitives, (2) comprehensive protection against differential cryptanalysis and chosen-plaintext attacks, and (3) built-in compliance with HIPAA/GDPR regulatory requirements. 
% The remainder of this paper is systematically organized to present our contributions: Section \ref{lit} critically analyzes related work in IoMT security, Section \ref{method} elaborates on SLIE's novel methodology and formal threat model, Section \ref{exp} validates our approach through extensive experimental results, Section \ref{discuss} provides a nuanced discussion of findings and practical limitations, and Section \ref{conclu} concludes our contributions.

\section{Literature Review}
\label{lit}
The sensitive nature of healthcare data makes it a prime target for cyberattacks, with breaches risking patient privacy, safety, and institutional reputation. Securing this data requires efficient, many-to-many, end-to-end encryption across low-power IoT devices. An ideal cryptographic framework must therefore provide robust identity authentication, key management, and access control while remaining adaptable to fast-paced emergency scenarios.

ABE presents a promising solution by integrating access policies into encryption, allowing decryption only when user attributes satisfy predefined criteria \cite{a19}. This approach simplifies key management compared to traditional PKI and enables granular access control. For instance, Satheesh and Sree \cite{a24} utilized bilinear mapping in ABE for secure decentralized cloud storage, Zhang et al. \cite{a26} applied Ciphertext-Policy ABE to streamline healthcare data encryption, and Obiri et al. \cite{a27} adopted Key-Policy ABE (KP-ABE) to enable flexible policy adjustments. However, ABE-based systems often rely on computationally intensive bilinear pairing operations, rendering them less suitable for dynamic, resource-constrained IoT environments.

As an alternative, IBE simplifies key distribution by using a user's identity (e.g., an email) as their public key, eliminating the need for pre-shared certificates \cite{a30}. Li et al. \cite{a34} further enhanced IBE with wildcard key derivation for improved access control and integrated Blockchain for decentralized authorization. While such approaches reduce overhead, they often retain a dependence on a CA, introducing a potential security bottleneck. Conversely, while lightweight cryptographic algorithms are designed for resource-constrained devices, they often lack end-to-end security structures, making them vulnerable to attacks.

Reflecting on these trade-offs, it is clear that the expansion of healthcare IoT demands a cryptographic framework that simultaneously addresses three critical dimensions for cross-device data sharing: (1) strong security guarantees, (2) efficient resource utilization, and (3) compliance with domain-specific policies.

\section{Methodology}
\label{method}
\subsection{Hierarchical Access Control with Time-Bound Encryption}
\label{revocation}
\begin{figure}[!htb]
    \centering
    \begin{minipage}{0.49\textwidth}
         \centering
    \includegraphics[width=\textwidth]{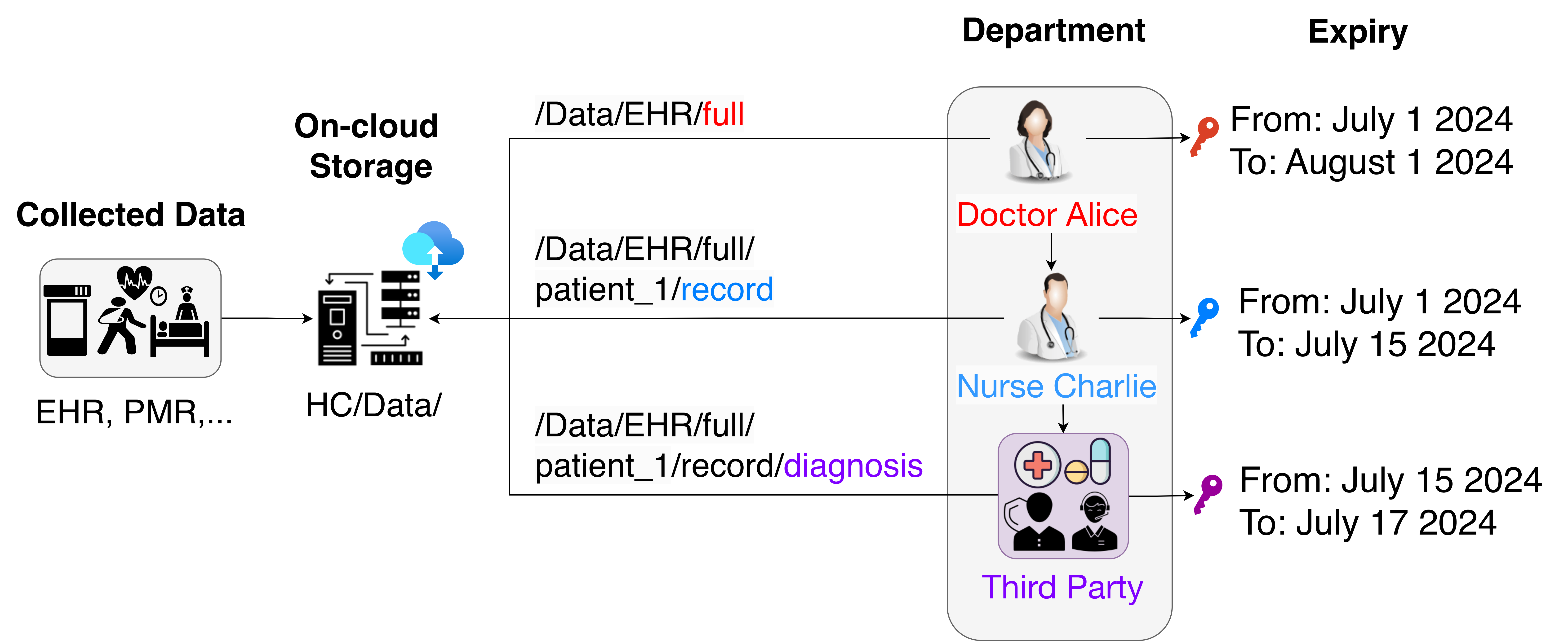}
    \caption{Hierarchical structure of healthcare communication}
    \label{expp}
    \end{minipage}\hfill
    \begin{minipage}{0.49\textwidth}
       \centering
    \includegraphics[width=\textwidth]{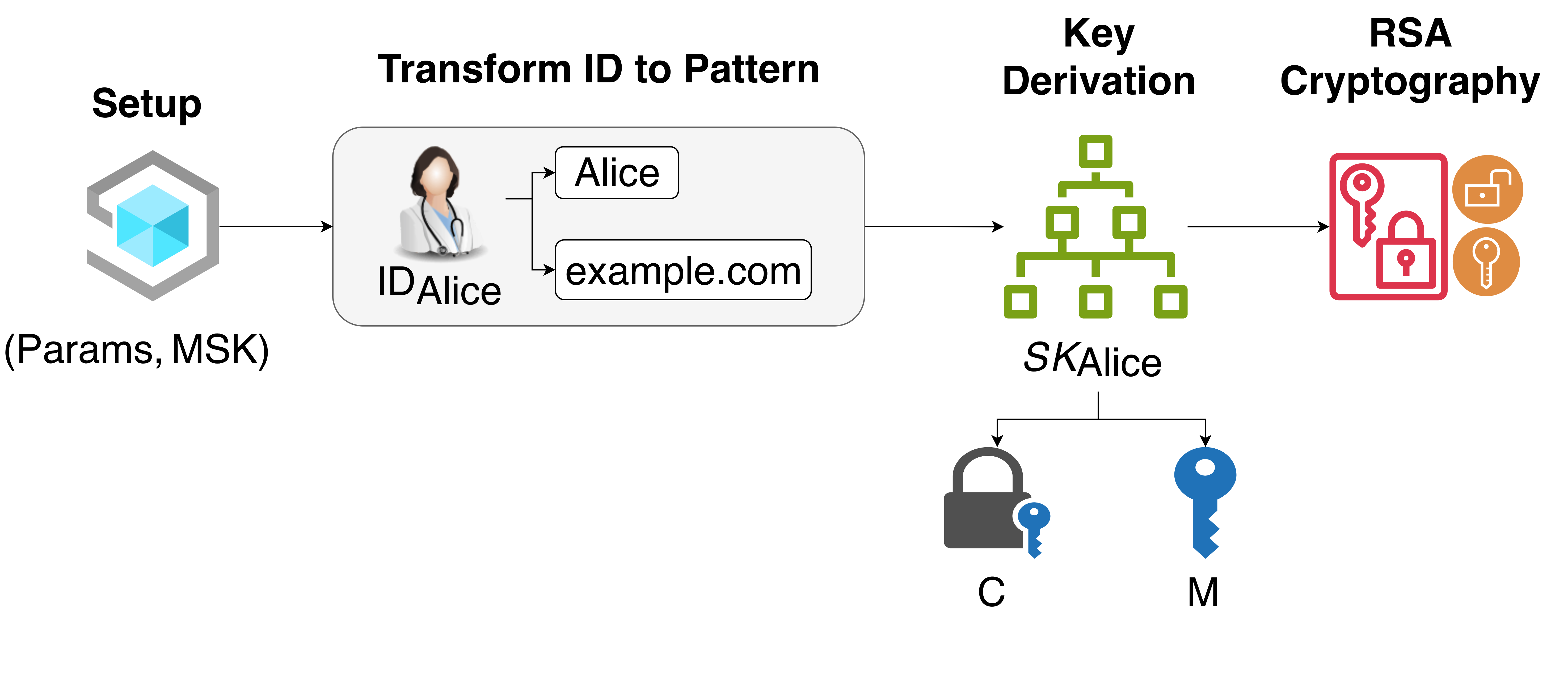}
    \caption{IBE-based key generation (RSA for double protection is optional)}
    \label{exp_ibe}
    \end{minipage}
\end{figure}

Our proposed framework integrates lightweight IBE with an efficient revocation mechanism, specifically designed for resource-constrained IoT devices handling sensitive cross-sectional healthcare data. Unlike existing IBE schemes that focus on general access control, SLIE addresses the unique requirements of healthcare hierarchies through role-specific temporal constraints and automatic delegation patterns that align with clinical workflows. Particularly, in Fig. \ref{expp}:

\begin{itemize}
\item Doctor Alice can access \texttt{/HC/Data/EHR/full} for entire departmental records
\item Nurse Charlie is restricted to specific patient records (\texttt{patient\_1/record})
\item Third-party providers only receive access to diagnostic outcomes (\texttt{diagnosis})
\end{itemize}
All data undergoes end-to-end encryption before cloud upload, eliminating plaintext exposure risks. The system implements temporally-graded access, with key validity periods scaled by role sensitivity  to balance security and operational flexibility in time-sensitive cases (e.g., stroke, myocardial infarction).

Keys have a validity period \( \tau \) that is automatically determined based on healthcare role requirements. This approach addresses a critical limitation in existing IBE schemes, where revocation typically requires maintaining centralized blacklists or complex re-encryption procedures. Our time-bounded approach eliminates these overheads through automatic expiration mechanisms. A key is considered valid or expired based on the current time \( t \)~\eqref{eq:expire}:

\begin{equation}
\label{eq:expire}
\text{Expire}(SK_{\text{Alice}}, t) = 
\begin{cases} 
\text{Valid}   & \text{if } t \leq \tau \\
\text{Invalid} & \text{if } t > \tau    
\end{cases}
\end{equation}

The expiration or validity period ($\tau$) for data access should be set through a risk assessment considering data sensitivity, role changes, regulations (e.g., HIPAA, GDPR), and institutional policies, as current regulations lack specific timeframes for healthcare roles. Particularly, in the U.S., with an average hospital stay of 5.5 days \cite{x}, nurses providing direct care should have a $\tau$ of 6–7 days, doctors expecting follow-ups a $\tau$ of up to 21 days, and third-party contractors (e.g., IT, medical services), with less 24/7 monitoring responsibility, a $\tau$ of up to 2 days. Family members may receive 1-day portal access. Nevertheless, this highly depends on the aforementioned risk assessment factors. These durations reflect the varying levels of access privilege and security risk associated with different roles. Specifically, the key lifecycle consists of the following steps:

\begin{enumerate}
    \item \textbf{Initial Key Assignment}: Alice receives a private key \( SK_{\text{Alice}} \) with an expiry time \( \tau_0 \).
    
    \item \textbf{Periodic Key Refresh}: When \( \tau_0 \) approaches, a new key \( SK'_{\text{Alice}} \) is issued with an extended expiry \( \tau_1 > \tau_0 \)~\eqref{eq:refresh}:
    
    \begin{equation}
    \label{eq:refresh}
    \text{Refresh}(SK_{\text{Alice}}, \tau_0) = (SK'_{\text{Alice}}, \tau_1)
    \end{equation}
    
    \item \textbf{Revocation by Non-Renewal}: If no new key is issued after expiry, Alice’s access is revoked by default.
\end{enumerate}

% Access keys for individuals (e.g., doctors, nurses, contractors) are renewed based on verified active employment, ongoing patient care responsibilities, incomplete contracted tasks, or extended patient-authorized hospital stays, as confirmed by hospital information systems prior to privilege assignment. Hence, this automated revocation mechanism obviates centralized revocation infrastructure, minimizes administrative overhead, ensuring immediate access termination upon role changes, and scales with organizational complexity.
Access keys for staff are automatically renewed based on active employment and patient responsibilities, as verified by hospital systems. This automated revocation eliminates centralized infrastructure, ensures immediate access termination, and scales with organizational complexity.

\subsection{Identity-Based Key Generation}
\label{ibe}
SLIE leverages WKD-IBE ~\cite{intro19} for key management (Fig. \ref{exp_ibe}), but extends the standard approach through healthcare-specific pattern designs over \((\mathbb{Z}_p^* \cup \{\bot\})^\ell\). Our contribution lies in the development of patterns that encode clinical hierarchies and role relationships, allowing fine-grained, non-commutative delegation and efficient key derivation across multiple concurrent healthcare departments encoded as disjoint slot partitions. An initial setup entity generates the parameters and a master secret key (MSK) \eqref{eq:setup}. For Alice, with identity \( ID_{\text{Alice}} \), a private key is derived \eqref{eq:keyder}. Messages are encrypted \eqref{eq:encrypt} and decrypted using Alice's private key \eqref{eq:decrypt}. In addition, keys are preserved on serve database. 

\begin{equation}
\label{eq:setup}
(\text{Params}, \text{MSK}) = \text{Setup}(1^\kappa)
\end{equation}
\begin{equation}
\label{eq:keyder}
SK_{\text{Alice}} = \text{KeyDer}(\text{Params}, \text{MSK}, \text{Pattern}(ID_{\text{Alice}}))
\end{equation}
\begin{equation}
\label{eq:encrypt}
C = \text{Encrypt}(\text{Params}, \text{Pattern}(ID_{\text{Alice}}), M)
\end{equation}
\begin{equation}
\label{eq:decrypt}
M = \text{Decrypt}(\text{Params}, SK_{\text{Alice}}, C)
\end{equation}

% The integration of healthcare workflow patterns into the WKD-IBE framework is distinctive. While traditional IBE schemes generate keys based on generic identity strings, SLIE's pattern-based approach encodes both role hierarchy and temporal constraints within the cryptographic structure itself, enabling automatic policy enforcement without requiring external access control mechanisms.
Unlike traditional IBE, which uses generic identity strings, SLIE's WKD-IBE framework integrates healthcare workflow patterns directly into the cryptographic structure. This encodes role hierarchy and temporal constraints, enabling automatic policy enforcement without external mechanisms.

\section{Experiments and Results}
\label{exp}
The SLIE evaluation was conducted on the LifeSnaps Fitbit Dataset \cite{snap}, a longitudinal repository containing over 71 million multi-modal health records from 71 participants collected over four months. The dataset's capture of 35+ clinical biomarkers at varying temporal resolutions (seconds to daily) provides a realistic benchmark for validating the security of healthcare IoT systems.

 % through Fitbit Sense wearables, ecological momentary assessments, and validated surveys, making it particularly suitable for security validation in digital health applications.

% Our experimental pipeline involved two key components: (1) A Joint Encryption and Delegation Infrastructure (JEDI) server implemented as a Docker container (built from the \texttt{go-jedi} directory) that handled dataset preprocessing and directory mounting, and (2) An Ethereum-based smart contract testing environment (implemented in the \texttt{kyc-contract} directory using Hardhat) for medical record validation. The test framework can be customized by modifying the \texttt{MedicalRecord-test} configuration file, enabling evaluation of SLIE's security properties under diverse healthcare data scenarios.

\subsection{Performance on Diverse Resource-constrained Devices}
Table \ref{tab:low-power-evaluate} illustrates the compatibility and performance of SLIE across a diverse range of resource-constrained healthcare IoT devices. It demonstrates SLIE’s adaptability to real-world healthcare applications by evaluating key operation, such as key generation, encryption, decryption, and delegation, which are averaged over 10 runs to ensure statistical reliability. Resource-rich devices like the MacBook Pro (2016) demonstrate exceptional efficiency, with encryption and decryption times of \SI{0.056}{\milli\second} and \SI{0.0755}{\milli\second}, respectively, and a notably low delegation time of \SI{14}{\milli\second}. In contrast, ultra-low-power devices such as the Apple Watch Series 6 achieve encryption and decryption in \SI{0.1}{\milli\second} and \SI{0.03}{\milli\second}, respectively, showcasing SLIE’s optimization for constrained environments. The Raspberry Pi 4 strikes a balance, with encryption and decryption times of \SI{0.3}{\milli\second} and \SI{0.4}{\milli\second}, suitable for edge computing in IoMT networks.

\begin{table}[H]
\centering
\caption{Performance evaluation of SLIE on various devices}
\label{tab:low-power-evaluate}
\resizebox{\textwidth}{!}{%
\begin{tabular}{|l|c|c|c|c|c|}
\hline
\textbf{Devices} & \textbf{Generate key} & \textbf{Encrypt} & \textbf{Decrypt} & \textbf{Delegate} \\ \hline
% Apple Watch Series 4 & 314ms & 56ms & 42ms & 144ms \\ \hline
% Apple Watch SE & 122ms & 0.5ms & 0.1ms & 83ms \\ \hline
Apple Watch Series 6  & 122ms & 0.1ms & 0.03ms & 82ms \\ \hline
Fitbit Versa 2 & 182ms & 42ms & 53ms & 83ms \\ \hline
Raspberry Pi 4 & 122ms & 0.3ms & 0.4ms & 82ms \\ \hline
Macbook Pro 2016 MLH42 & 99ms & 0.056ms & 0.0755ms & 14ms \\ \hline
\end{tabular}%
}
\end{table}

\Cref{tab:power-consumption} details the power consumption and resource utilization of SLIE’s core operations on representative IoT devices. Encryption and decryption operations exhibit remarkable energy efficiency, consuming \SI{0.014}{\joule} and \SI{0.009}{\joule} per \SI{1}{KB} data respectively. These metrics highlight SLIE’s suitability for battery-constrained devices, enabling prolonged operation in wearable and implantable medical devices. The delegation operation, critical for secure access control in IoMT systems, consumes \SI{0.052}{\joule}/KB. While delegation is more resource-intensive, its moderate power usage (\SI{1.45}{\watt}) and CPU utilization (\SI{35.6}{\percent}) ensure feasibility on low-power platforms. Memory usage remains modest, ranging from \SI{12.5}{\mega\byte} (decryption) to \SI{22.3}{\mega\byte} (delegation), further supporting SLIE’s lightweight design. Collectively, these attributes demonstrate SLIE as an energy-efficient and scalable solution for secure data sharing for IoMT services, balancing performance, power, and security requirements.

\begin{table}[!htb]
\centering
\caption{Power consumption analysis of SLIE operations}
\label{tab:power-consumption}
\resizebox{\textwidth}{!}{%
\begin{tabular}{|l|c|c|c|c|c|}
\hline
\textbf{Operation} & \textbf{Exec. time (ms)} & \textbf{Power Usage (W)} & \textbf{Mem. Usage (MB)} & \textbf{CPU Util. (\%)} & \textbf{Energy Eff. (J/KB)}  \\ \hline
Encrypt & 12 & 1.20 & 16.80 & 27.50 & 0.014 \\ \hline
Decrypt & 9 & 0.95 & 12.50 & 19.20 & 0.009 \\ \hline
Delegate & 18 & 1.45 & 22.30 & 35.60 & 0.052 \\ \hline
\end{tabular}%
}
\end{table}

\subsection{Scalability and Efficiency in IoT Data Encryption}
% Interpreting scalability and performance
\Cref{tab:SLIE-performance} quantifies SLIE’s performance across data sizes from \SI{1}{\kilo\byte} to \SI{10}{\mega\byte}, averaged over 10 runs. For smaller data sizes (\SI{1}KB), SLIE achieves encryption and decryption times of \SI{0.936}{\milli\second} and \SI{0.217}{\milli\second}, respectively, with minimal CPU usage (\SI{0.5}{\percent} and \SI{0.7}{\percent}) and RAM consumption (\SI{1520}{\kilo\byte} and \SI{1300}{\kilo\byte}). As data sizes increase, SLIE metrics increase linearly, demonstrating its suitability for resource-constrained IoMT devices, such as wearables and edge nodes.

% to \SI{10}{\mega\byte}, encryption and decryption times scale to \SI{25.036}{\milli\second} and \SI{15.119}{\milli\second}, with CPU usage rising to \SI{8.3}{\percent} and \SI{9.2}{\percent}, and RAM usage peaking to \SI{55120}{\kilo\byte} and \SI{50800}{\kilo\byte}. This linear scalability underscores SLIE’s suitability for resource-constrained IoMT devices, such as wearables and edge nodes.

\begin{table*}[!htb]
\centering
\caption{Performance metrics of SLIE under different scenarios}
\label{tab:SLIE-performance}
\resizebox{\textwidth}{!}{%
\begin{tabular}{|c|c|c|c|c|c|c|c|}
\hline
\textbf{Data Size} & \makecell{\textbf{Key Creation}\\\textbf{Time (ms)}} & \makecell{\textbf{Encryption}\\\textbf{Time (ms)}} & \makecell{\textbf{Encryption}\\\textbf{CPU Usage (\%)}} & \makecell{\textbf{Encryption}\\\textbf{RAM Usage (KB)}} & \makecell{\textbf{Decryption}\\\textbf{Time (ms)}} & \makecell{\textbf{Decryption}\\\textbf{CPU Usage (\%)}} & \makecell{\textbf{Decryption}\\\textbf{RAM Usage (KB)}} \\ \hline
1KB  & 81  & 0.936  & 0.5  & 1520  & 0.217  & 0.7  & 1300  \\ \hline
% 10KB & 83  & 1.482  & 1.2  & 2340  & 0.477  & 1.5  & 2200  \\ \hline
100KB & 81  & 3.500  & 1.8  & 5200  & 1.184  & 1.9  & 4500  \\ \hline
% 250KB & 81  & 3.981  & 2.1  & 7120  & 1.391  & 2.4  & 6900  \\ \hline
500KB & 85  & 4.769  & 2.8  & 9650  & 1.846  & 3.0  & 8800  \\ \hline
% 750KB & 81  & 8.935  & 3.6  & 11320 & 2.100  & 3.8  & 10900 \\ \hline
1MB   & 83  & 12.677 & 3.2  & 12000 & 3.318  & 4.2  & 11500 \\ \hline
5MB   & 95  & 15.530 & 4.5  & 35600 & 8.993  & 5.5  & 30400 \\ \hline
% 7MB   & 70  & 20.429 & 6.1  & 42340 & 10.270 & 6.8  & 41200 \\ \hline
10MB  & 83  & 25.036 & 8.3  & 55120 & 15.119 & 9.2  & 50800 \\ \hline
\end{tabular}%
}
\end{table*}

% Comparing with benchmarks
% As benchmarked in \Cref{tab:comparison-metrics}, SLIE demonstrates a compelling performance and security profile compared to existing schemes. It significantly outperforms RSA, which recorded encryption, decryption, and key generation times of \SI{121.85}{\milli\second}, \SI{62.93}{\milli\second}, and \SI{1476.57}{\milli\second}, respectively. It means an 84.54\% improvement in encryption speed, a 99.70\% improvement in decryption speed. While SLIE's raw speed is lower than the symmetric cipher ChaCha20 (\SI{3.57}{\milli\second} encryption, \SI{2.81}{\milli\second} decryption), it provides critical security features that ChaCha20 lacks. These empirical results position SLIE as a more optimal solution for IoMT services.

As benchmarked in \Cref{tab:comparison-metrics}, SLIE demonstrates a compelling balance of performance and security. It drastically outperforms RSA, achieving an 84.54\% faster encryption and a 99.70\% faster decryption. While its raw speed is lower than the symmetric cipher ChaCha20, SLIE provides essential, built-in security features—such as hierarchical access control—that ChaCha20 lacks. These suggest that SLIE a more optimal solution for IoMT services.

\begin{table}[!htb]
\centering
\caption{Average performance metrics comparison}
\label{tab:comparison-metrics}
\resizebox{\textwidth}{!}{
\begin{tabular}{|c|c|c|c|c|c|}
\hline
\textbf{Algorithm} & \textbf{Key Creation (ms)} & \textbf{Enc (ms)} & \textbf{Dec (ms)} & \textbf{CPU Usage (\%)} & \textbf{RAM Usage (KB)} \\ \hline
SLIE & 82.3 & 9.73 & 4.49 & 3.66 & 17,535 \\ \hline
RSA & 121.85 & 62.93 & 1476.57 & 5.86 & 31,388 \\ \hline
ChaCha20 & None & 3.57 & 2.81 & 2.33 & 14,775 \\ \hline
\end{tabular}}
\end{table}

\section{Conclusion}
\label{conclu}
% SLIE, a lightweight cryptosystem utilizing WKD-IBE, ensures secure, scalable, and efficient omnidirectional communication for IoMT services. Performance evaluations show SLIE outperforms RSA by 84.54\% in encryption time and 99.70\% in decryption time, with 2.6\% time increase per 1MB, addressing ChaCha20’s lack of hierarchical access control, automatic revocation, and key generation. With energy efficiency of 0.014J/KB for encryption and \SI{0.0009}{\joule/KB} for decryption, SLIE suits resource-constrained devices (e.g., Apple Watch Series 6, Raspberry Pi 4) (\Cref{tab:low-power-evaluate}). Hierarchical access control and expiry-based key revocation ensure HIPAA/GDPR compliance.  Future work will integrate post-quantum primitives, optimize for implantable devices, and validate interoperability with healthcare standards (e.g., FHIR) through large-scale hospital deployments.
In conclusion, the proposed SLIE cryptosystem establishes a robust foundation for secure IoMT services by combining the efficiency of a lightweight WKD-IBE framework with scalable trust management. It achieves a compelling performance profile, surpassing RSA in speed by over 84\% for encryption and 99\% for decryption while operating with high energy efficiency (0.014 J/KB). Crucially, SLIE overcomes the key management shortcomings of symmetric systems like ChaCha20 by natively supporting hierarchical access control and automatic key revocation, ensuring both security and regulatory compliance. Looking ahead, our research will evolve to incorporate post-quantum security, optimize for next-generation implantables, and demonstrate real-world interoperability with standards such as FHIR in clinical settings.

\section{Code Availability}
The source code developed for this study and the scripts used to generate the results can be found at \url{https://github.com/SonHaXuan/SLIE}.

\section*{Acknowledgement}
The authors acknowledge Ho Chi Minh City University of Technology (HCMUT), VNU-HCM, for supporting this study.

% Security tests confirm 0\% MITM, $<5\%$ side-channel, and $<5\%$ power analysis success (95\% CI: [0.00, 0.08], p $< 0.001$, Cohen’s d $\geq 0.85$), using constant-time operations, memory obfuscation, and DBDH assumption (\Cref{tab:security_results}). SLIE’s O(log n) complexity via HIBE outperforms DAC-MACS (O(n·m)) and others (O(n)) (\Cref{tab:security_comparison}). NIST-aligned hybrid encryption provides post-quantum readiness, and memory sanitization (\texttt{runtime.MemStats}) prevents data leakage. Despite moderate quantum resistance, SLIE’s decentralized key management and linear scalability is transformative for IoMT. Future work will integrate post-quantum primitives, optimize for implantable devices, and validate interoperability with healthcare standards (e.g., FHIR) through large-scale hospital deployments.

% \newpage
\bibliographystyle{unsrt}
% \bibliography{ref}

\newpage
\end{document}